# Cerebellar Contributions to Action and Cognition: Prediction, Timescale, and Continuity

Jonathan S. Tsay[1] & Richard B. Ivry[2]

[1]Department of Psychology, Carnegie Mellon University
[2]Department of Psychology, University of California, Berkeley

## Abstract

The cerebellum is implicated in nearly every domain of human cognition, yet our understanding of how this subcortical structure contributes to cognition remains elusive. Efforts on this front have tended to fall into one of two camps. On one side are those who seek to identify a universal cerebellar transform, a single algorithm that can be applied across domains as diverse as sensorimotor learning, social cognition, and decision making. On the other side are those who focus on functional specializations tailored for different task domains. In this perspective, we propose an integrated approach, one that recognizes functional specialization across different cerebellar subregions, but also builds on common constraints that help define the conditions that engage the cerebellum. Drawing on recurring principles from the cerebellum's well-established role in motor control, we identify three core constraints: (1) Prediction—the cerebellum performs anticipatory, not reactive, computations; (2) Timescale—the cerebellum generates predictions limited to short intervals; and (3) Continuity—the cerebellum transforms continuous representations such as space and time. Together, these constraints define the boundary conditions underlying when and how the cerebellum supports cognition, and, just as importantly, specify the types of computations that should not depend on the cerebellum.

## I. The Cognitive Cerebellum

In the early years of the cognitive neuroscience revolution, Leiner, Leiner, and Dow offered the conjecture that the cerebellum helps coordinate mental processing, similar to how it helps coordinate movement (1). Their observations were mostly based on gross anatomy and evolutionary considerations, and in particular, what they saw as the parallel expansion of the prefrontal cortex and lateral cerebellum. Since then, a diverse literature has emerged over the past 40 years implicating the cerebellum in a broad array of cognitive, social, and affective function (2–5). This work has led to the establishment of a new clinical diagnostic category, Cerebellar Cognitive Affective Syndrome (CCAS) (6–8), in recognition of the symptoms that can be observed in patients with neurological disorders of the cerebellum. Complementing the work involving human participants, there has been a surge in studies with non-human species to explore the circuitry and physiological processes that underlie the contribution of the cerebellum to non-motor behavior such as spatial cognition, social cognition, and reinforcement learning (9–11).

To date, much of this work has been descriptive, focusing on the question of "Is the cerebellum involved in X" where X might be language, attention, working memory, social cognition, emotional regulation, etc. Similarly, the items on the CCAS battery have been selected because of their diagnostic value in evaluating impairments in these domains (6). Studies using functional neuroimaging or neurophysiological recordings have probed how information is encoded in the cerebellum, with the results providing detailed maps depicting functional specialization in non-motor domains across much of the cerebellar cortex (Figure 1A) (5, 12, 13).

While this work has certainly led to an appreciation of the broad functional domain of the cerebellum, the field still struggles to understand *how* the cerebellum contributes to cognition. We have a large catalogue showing that the cerebellum is engaged in a broad range of tasks but have made limited progress in deepening our mechanistic understanding, be it models that describe computational principles at the psychological level or detail how these computations emerge from the physiological activity of the cerebellum.

In seeking to understand how the cerebellum supports cognition, many researchers have drawn inspiration from the idea that computations supporting motor control might extend to cognition. This approach is motivated by the strikingly uniform architecture of the cerebellar cortex (14, 15). Although there are variations across subregions (16), the basic cell types and connectivity patterns have been maintained over the long evolutionary history of the cerebellum. A second motivation comes from the fact that this architecture has inspired sophisticated models of how the cerebellum supports sensorimotor coordination and learning. While these models have and continue to undergo refinement, core computational principles have been co-opted into hypotheses explaining how the cerebellum supports mental coordination and non-motor learning.

These principles are captured by what is commonly referred to as the Marr-Albus-Ito model. Impressed by the massive degree of divergence that arises from the parallel fibers, the axons of the 50 billion granule cells that provide the main input to the cerebellar cortex, Marr proposed that the cerebellum operates as a pattern recognition system (17). Focusing on sensorimotor control, pattern recognition would be essential for adaptive behavior, specifying the context that enables organisms to produce a near-infinite set of movements (18, 19). For example, detecting whether one is walking on ice or solid ground is critical for adjusting movement kinematics to prevent a fall (20).

Building on Marr's theory, Albus and Ito proposed a learning scheme in which the climbing fibers, the second input to the cerebellar cortex, provide error information to shape how the Purkinje cells respond to the parallel fiber input (21, 22). Contextual information, including an *efference copy* of descending motor commands, can be used to *predict* the sensory consequences of an upcoming action—an operation later termed a *forward model* (Figure 1B) (23). These predictions modulate downstream motor circuits, allowing movements to be rapidly coordinated without relying on delayed sensory feedback. Interneurons within the cerebellar cortex refine the *timing* of these predictions, ensuring they influence behavior precisely when needed (24, 25). When the sensory outcome deviates from the predicted outcome, the activation of the climbing fibers serves as an *error signal*, triggering corrective actions as well as inducing plasticity to update the forward model. This error-driven process incrementally refines behavior, helping ensure that the movements are accurate and performed in a *continuous*, coordinated manner (26).

Extending the principles of the Marr-Albus-Ito model to cognition has led to the hypothesis that the cerebellum builds forward models that run internal simulations to generate domain-general predictions (3, 23, 27, 28). For example, in the language domain, semantic representations in the cerebellum could be used to generate predictions regarding the upcoming words in an utterance (29–31). In the social domain, the cerebellum might generate a forward model of the internal state of another agent and anticipate their behavior during a social interaction (10, 32).

While the generalized prediction hypothesis offers a compelling starting point, the mapping of core principles between motor and cognition is not straightforward. Consider the learning process: In motor control, the cerebellum receives an efference copy of the motor command and, in combination with inputs specifying the current context, generates a prediction of the expected consequences of that movement. Learning within this system is driven by sensory prediction errors, the difference between the expected and actual sensory outcome. These errors have a clear vector, which, through gradient descent, can be used to guide corrections that progressively improve the action (see reviews: (33, 34)).

Extending this idea to social cognition, the input can be viewed as the current context, including the state of other agents. A forward model could be used to generate predictions, in this case, anticipating the behavior of other individuals. A mismatch between the expected and actual outcome could define a social prediction error. Yet, here, the analogy faulters: What is the "vector" of a social prediction error? How does it specify the direction of change, and toward what target? Without a defined structure or goal, it remains unclear how social prediction errors could guide learning.

A broader concern with the generalized prediction hypothesis is that predictive processing is not unique to the cerebellum. Indeed, as exemplified in the Predictive Coding Framework, prediction is viewed as the canonical computation of the brain (35), one that is performed at all stages of neural activity. As such, if prediction is the hallmark of all neural processing, the key question then becomes: What is unique about cerebellar-dependent predictions?

The concept of prediction when used to describe non-motor functions of the cerebellum is often invoked in an underspecified manner. For instance, it is reasonable to suppose that the ability to anticipate future states will enhance "mental coordination", supporting cognitive processes across domains (2). Yet without clearly defined constraints, such claims risk becoming too broad for rigorous testing. What would disprove such a hypothesis? What are the cognitive symptoms of disrupted cerebellar predictions, and under what conditions do they emerge?

Equally important, *specifying* the constraints on cerebellar-dependent prediction should yield not only predictions about tasks impaired by cerebellar pathology or altered by cerebellar stimulation, but also about those that remain unaffected. In the rush to define the cerebellum's role in cognition, spared functions are often overlooked— yet they are diagnostic of what makes cerebellar processing unique.

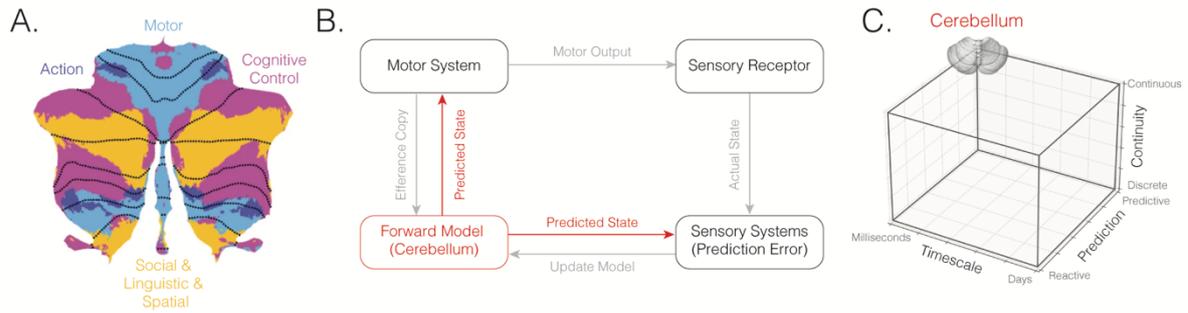

**Figure 1. Constraints Governing Cerebellar Computation in Action and Cognition. A)** Functional map of the cerebellum, divided into regions linked to movement (activated for body-part–specific movements), action (action observation and motor imagery), cognitive control, and social-linguistic-spatial processes (adapted from Nettekoven et al., 2024). **B)** The cerebellum as a forward model, using an efference copy of the motor command to predict the sensory consequences of the movement. These state predictions can guide movement, compensating for processing delays during motor control. Sensory prediction errors, the difference between predicted and actual outcomes, may be computed throughout the system and arrive at the cerebellum via a comparator (e.g., inferior olive). By analogy, the cognitive cerebellum may predict changes in perceptual or mental states, relay these predictions to association areas in cortex, and incorporate error feedback to update the forward model for cognitive control. **C)** 3D space highlighting that the cerebellum is essential for short-timescale predictions that involve continuous transformations in space and time, which we term continuous representational transformations (CoRT).

## II. Constraints Governing Cerebellar Computation in Action and Cognition

What constraints govern cerebellar computation in action and cognition? To tackle this question, we consider the role of the cerebellum in motor control, identifying recurring principles that have emerged over decades of research. We use these constraints as a springboard to ask whether similar principles extend to cerebellar computation in a range of task domains.

Because sensory feedback from movement is delayed, the nervous system cannot operate as a closed-loop system for real-time motor control. To avoid the instabilities such delays would cause, the sensorimotor system must anticipate future states, ensuring a smooth, continuous transition from the current state to the desired state (23). As described in Section I, the cerebellum is a core node in this control scheme, generating predictions about the sensory information that will arise from these upcoming state transitions (i.e., a forward model). To be effective, these predictions must integrate both time and space: Reaching for a water glass requires anticipating the sensory contact at a specific limb position and at a precise moment in time.

This model provides an elegant account of key features of cerebellar ataxia (CA). These individuals struggle to anticipate the kinematic and dynamic changes that occur during movement (36, 37). Unlike patients with apraxia or hemiplegia, they can recruit the necessary muscles and show little loss of force. However, impaired feedforward control prevents proper coordination of agonist–antagonist activation, making it difficult to stop the movement at the intended position and time. As a result, their movements are ataxic—undershooting or overshooting the target—and, with increased reliance on delayed sensory feedback, exhibit intention tremor (38, 39). Transient disruption of cerebellar activity, either through transcranial magnetic stimulation in humans (Figure 2A) (40) or nerve stimulation/optogenetics in animal models (41, 42) produces similar ataxic errors, reinforcing the cerebellum's role in generating predictions of the limb's state across space and time.

The importance of the cerebellum for predictive, feedforward control—and its relative unimportance for reactive, feedback control—is evident across a wide range of sensorimotor tasks (43). One of the best studied is sensorimotor adaptation, in which an unexpected perturbation is introduced during movement (Figure 2B). The resultant sensory prediction error is used to recalibrate the sensorimotor system and, correspondingly, refine the forward model associated with that motor command (33, 44–

46). Individuals with CA show impairments on adaptation tasks across a wide range of effectors and tasks (36, 47–52).

In contrast, individuals with CA do not exhibit impairments on tasks that depend on reactive, feedback-driven control (43). Indirect evidence comes from the preservation of slow movements, where the need for predictive braking from the antagonist muscle is minimal (49). More direct evidence of intact closed-loop control comes from studies that look at rapid mid-movement corrections to perturbed feedback signals. For example, when a visual target is displaced during a reach, individuals with CA exhibit similar corrective adjustments as controls (53–55).

The dissociation between impaired feedforward and preserved feedback mechanisms extends beyond motor control into perception. Individuals with cerebellar ataxia (CA) often misjudge limb position after rapid goal-directed movements, when localization relies on sensory predictions from a forward model; in contrast, they localize accurately after passive movements, which depend solely on proprioceptive feedback (56). Indeed, this dissociation is observed across multiple sensory domains (57).

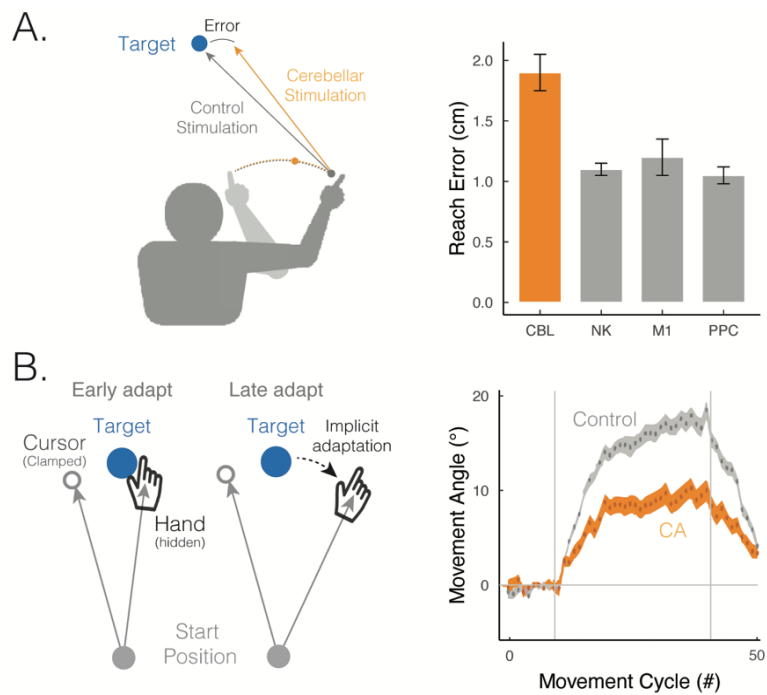

Figure 2. The Cerebellum as a Forward Model in Motor Control and Motor Learning. A) Online use of predictive control. Participants moved their hand laterally (dotted line) until a tone cued a reach to the target (solid line; **left**). The error observed after cerebellar TMS suggests that planning is made from the actual hand position at tone onset (yellow dot) rather than the predicted future position the hand will reach (grey dot) due to delays in processing the sensory feedback and updating the motor output (adapted from Miall et al., 2007). This effect was not observed following TMS stimulation over M1 or posterior parietal cortex (PPC), or a control site, the neck (NK) (**right**). B) Clamped feedback task to isolate cerebellar-dependent adaptation (**left**). When instructed to reach straight to a visual target while ignoring a cursor that always moves at a fixed angular offset from the target, participants exhibit an implicit shift in movement angle, akin to what is observed in standard visuomotor adaptation tasks. Implicit adaptation is attenuated in patients with cerebellar ataxia (CA; orange) compared to controls (gray) (**right**; unpublished data; also see (48, 58, 59)).

Drawing on this rich literature, we highlight three core constraints on cerebellar-dependent computations:

**Prediction Constraint: Cerebellar computations support feedforward, predictive control—not feedback-based, reactive responses**. The cerebellum will be engaged in tasks that require anticipating future states based on an internal transformation of a current representation. We assume that such

predictions involve a simulation process that estimates the transition from the current state to the desired state. The corollary here is that the cerebellum should *not* be involved in tasks where predictions can be based on retrieval processes, at least when there is no requirement to manipulate or transform internal representations. For example, predictions based on episodic memory can arise from recalling prior associations: Upon entering a favorite coffee shop, I can confidently predict my friend will order an espresso.

**Timescale Constraint: Cerebellar computations operate over short timescales—not extended temporal intervals**. Predictions extend over different time scales. A large body of literature shows that the cerebellum is critical for tasks that require high temporal precision, but only over a limited temporal range (milliseconds). Such precision is essentially for coordinating the muscular activation pattern that allows our fingers to grasp an object. The corollary here is that the cerebellum should *not* be involved in predictions that are temporally diffuse; for example, planning what we will make for dinner this evening.

**Continuity Constraint: Cerebellar computations perform state estimation over continuous dimensions of space and time—not discrete or categorical variables**. This constraint holds that the cerebellum is essential when predictions involve the transformation of a continuous variable. This encompasses movement where the corporeal body must move in a continuous manner, smoothly transitioning from one state to another. It would also apply to other contexts in which the prediction requires a continuous transformation such as anticipating the trajectory of a moving object. The corollary here is that the cerebellum should *not* be required for predictions involving discrete, symbolic, or categorical variables. For example, anticipating the final word of a sentence based on contextual association frequencies: "The man loosened his tie around his….neck."

Evaluated as a whole, these three constraints—grounded in the well-established motor control literature—suggest that the cerebellum is essential for computations involving **continuous representational transformations**, or what we refer to by the acronym **CoRT** (Figure 1C). In the following sections, we consider the relevance of these constraints for cognition, with an eye on sharpening our understanding of *how* the cerebellum might support mental coordination. Importantly, these constraints lend themselves to rigorous empirical tests, specifying which tasks require cerebellar computation and which do not.

## III. Empirical Support for CoRT

**Cerebellar Involvement in Continuous Temporal Prediction over Short Intervals**. Studies of temporal prediction, the ability to anticipate when an event will occur, have provided some of the clearest evidence for cerebellar-dependent computation (for reviews, see (3, 60)). A prominent example comes from the eyeblink conditioning literature (Figure 3A). In a basic form of this classical conditioning task, a tone precedes an air puff delivered to the eye. Initially, the animal reflexively blinks in response to the air puff, but with repeated pairings, the tone alone elicits the response. Richard Thompson and colleagues in the early 1980s reported the striking finding that lesions of the rabbit cerebellar cortex abolished the conditioned response but had minimal impact on the unconditioned response (61). This dissociation, since replicated in multiple species (62–65), underscores that the deficit is not motoric in nature, but rather reflects an impairment in associative learning (66).

The conditioned eyeblink is only adaptive if it is precisely timed with respect to the onset of the air puff, ensuring the eye is protected at the right moment. Interestingly, a conditioned response acquired in a similar paradigm, heart rate deceleration, remains unaffected by the same lesion (Figure 3B) (67). Heart rate deceleration unfolds over longer timescales and lacks temporal specificity, likely reflecting the engagement of the autonomic system in response to an aversive event.

How the cerebellum refines the timing of the conditioned eyeblink has been the focus of extensive research. A central idea emerging from this work is that interactions among interneurons of the cerebellar cortex generate a functional 'basis set' of timing elements. Metaphorically, these can be thought of as a collection of hourglass timers, tuned to activate the Purkinje cells at specific intervals (68) (see review: (69)). In this way, the representation of the tone is temporally maintained. In response to the error signal arising from the air puff, the synaptic activity of a subset of these elements will be depressed. This will reduce Purkinje cell inhibition of the deep cerebellar nuclei and thus, elicit a conditioned response that is timed to ensure the eye is protected at the onset of the air puff.

This simple associative learning paradigm highlights all three of our core constraints. Under the Prediction constraint, the cerebellum is essential for the predictive, conditioned eyeblink, but not essential for the reactive, non-predictive unconditioned response. Under the Timescale constraint, the cerebellar-dependent conditioned response is precisely timed, tailored to match the rapid inter-stimulus interval between the tone and air puff; the cerebellum is not required for the temporally diffuse, conditioned heart rate response when the interval extends beyond the millisecond range (63). Under the Continuity constraint, the basis set defines a continuous representation of time in the cerebellar cortex for precise temporal anticipation (70).

The cerebellum is also essential for temporal predictions in perceptual tasks (Figure 3C). Individuals with CA exhibit elevated discrimination threshold when comparing the duration of stimuli in the millisecond range (71), an effect that is also observed following non-invasive cerebellar brain stimulation in healthy individuals (72, 73). The specificity of this impairment is evident in findings showing preserved performance when the psychophysical judgment is based on non-temporal properties such as loudness (71). Neuropsychological and neuroimaging studies have associated the cerebellum with a bevy of other psychophysical tasks that require precise millisecond-level temporal resolution (74–77).

An intriguing dissociation comes from the work of Grube and colleagues (78). Across a set of time discrimination tasks, they found that CA participants were impaired when the target interval was presented in isolation, but not when the target interval was embedded in a rhythm. This pattern points to two distinct modes of timing: As with eyeblink conditioning, judging the duration of an isolated interval appears to rely on cerebellar-dependent temporal basis sets, a mechanism that provides a continuous representation of time. From a prediction perspective, the stimulus onset triggers the replay of an internalized standard interval, with the judgment based on whether the test stimulus ends before or after the standard's offset. In contrast, with rhythmic stimuli, the brain can anchor to the beat through entrainment, using each beat as a discrete point for comparison (did the test stimulus occur before or after a beat?).

The selective involvement of the cerebellum in interval-based timing has also been observed in attentional orienting tasks. Perceptual detection and discrimination can be enhanced when a cue indicates when a stimulus is likely to appear. CA participants show a marked deficit when the cue is a single interval (requiring continuous temporal replay), but perform normally when the cue is a series of beats (requiring a discrete comparison), even when task difficulty is matched (79, 80). Interestingly, patients with Parkinson's disease show the opposite pattern: They are impaired in using the rhythmic cues to orient attention but perform similarly to control participants with interval-based cues. Altogether, these findings strongly support the constraints of CoRT, underscoring the cerebellum's specialized role in generating continuous predictions over short intervals.

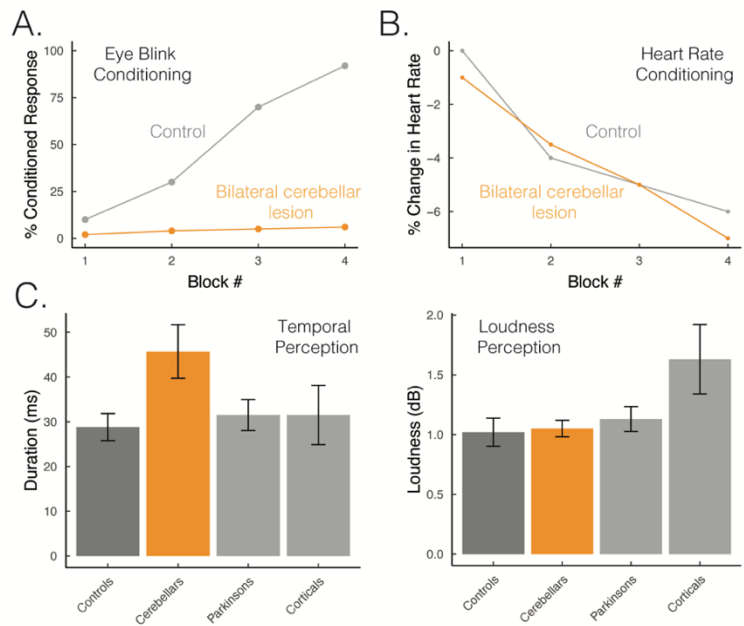

Figure 3. Cerebellar involvement in temporal prediction over short intervals. **A)** Rabbits with bilateral cerebellar lesions (orange) show abolition of eyeblink conditioning compared to controls (grey), measured as the percentage of conditioned responses after repeated pairing of a tone and air puff to the eye. **B)** In contrast, heart-rate conditioning, measured as conditioned heart rate deceleration when the tone is paired with a face shock, is spared (adapted from Lavond et al., 1984). **C)** Difference thresholds for perceptual judgments of duration **(left)** and loudness **(right)** across patient groups and age-matched controls. Cerebellar patients (orange) exhibit increased variability only on the duration task (adapted from Ivry & Keele, 1989).

**Cerebellar Involvement in Continuous Spatial Prediction over Short Intervals**. Coordinated movement requires anticipation of how the body will interact with the environment in time and space. As physical entities, the transformations required to plan and control a movement must be continuous, anticipating future states of the body and environment in a time-sensitive manner. In this section, we highlight non-motor tasks that rely on continuous spatial transformations and provide evidence that the integrity of the cerebellum is essential for optimal performance.

Our visual system has an incredible capacity for object recognition. Deciphering a complex scene is challenging, not only because it is cluttered with many objects, but also because factors such as occlusion or shadows can introduce ambiguities concerning surface ownership. Moreover, recognition may be made difficult if the to-be-perceived object is in an atypical orientation with respect to the viewer. In such situations, recognition is facilitated by mentally manipulating the image to align with what we consider its prototypical orientation (81).

Experimentally, this process has been studied with mental rotation tasks. In the classic work of Shepherd and colleagues, participants judged if a letter, when viewed upright, would be in its normal orientation or mirror reversed. The key finding was that reaction time increased in a near-linear manner as the angle of the stimulus relative to upright was increased (82). This observation, coupled with data from multiple measures, indicates that the internal transformations required on such tasks appear to be continuous: For example, when presented with a letter that is tilted by 30°, we cannot internally reformat the letter to its upright position without passing through the intervening orientations. This continuity constraint has been hypothesized to reflect the internalization, or embodiment, of how we might manipulate an object if inspecting it from multiple perspectives (83, 84).

To test whether the cerebellum supports continuous spatial transformations, we assessed individuals with CA on a variant of the mental rotation task (Figure 4A) (85). The CA group was slower than matched controls in judging if the target letter was in its normal orientation or mirror reversed. Importantly, this slowing became more pronounced with the magnitude of the required rotation. The

Group x Angle interaction suggests that the patients are slower in the rate at which they can internally manipulate the mental image. Notably, they did not exhibit elevated error rates. This pattern of results, normal accuracy but an increased slope in the RT function, is consistent with the hypothesis that the cerebellum facilitates an internal spatial transformation. Indeed, all CoRT constraints would appear to be operative during mental rotation: Over a short time scale, a continuous spatial transformation is applied to assess if the presented stimulus matches a predicted stimulus (e.g., normal orientation).

We used CoRT to guide our selection for a control task to compare with mental rotation. As prerequisites, we required the control task to involve visual matching and, critically, to demand iterative retrieval of discrete visual representations rather than continuous stimulus transformation. Given these considerations, we employed two spatial working memory control tasks ([Figure 4B](#)) (86). In each one, a series of objects of variable length (e.g., 1-5 items) was presented one at a time at different locations. After a short delay, a single object was shown, and the participant made a speeded response based on its spatial position (e.g., ordinal position in the time series of the probe item). Although RT increased with working memory load in both groups, the similar slopes for CA and control participants suggest the cerebellum is not involved in discrete memory retrieval.

The same dissociation was observed in a second experiment involving a different group of individuals with CA. Compared to control participants, the CA group exhibited a larger slope (i.e., slower rate) on the mental rotation task and a similar slope on a spatial working memory task, where participants judged whether a cued item was new or previously seen in the memorized list. These studies in the domain of visual cognition offer a mechanistic account of the CA groups' performance based on the Continuity constraint. These individuals were less facile in mentally rotating an internal visual image, an operation we assume requires the continuous transformation of an internal representation. In contrast, their rate of retrieval from spatial working memory was on par with the control group, presumably because the cerebellum is not essential for the discrete transformations required when iteratively searching through working memory.

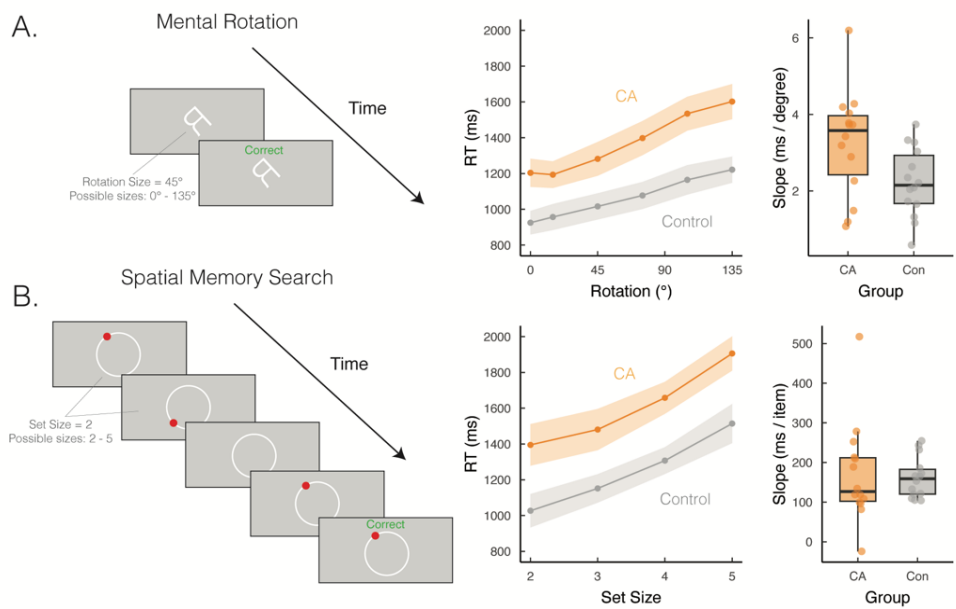

**Figure 4. Cerebellar involvement in continuous visual cognition. A)** Mental rotation task **(left)**. Participants judged whether a letter, if viewed upright, would be in its normal orientation or mirror reflected. Letters were presented at various orientations. Reaction time increased with the angular deviation of the stimulus **(middle)**. Patients with cerebellar ataxia showed a slower rate of rotation as inferred from the slope of this function **(right)**. **B)** Spatial working memory task **(left)**. Participants viewed a sequence of 2–5 circles, each shown for 1 s at random positions along a ring. After a 2 s delay, a probe appeared, and participants reported its ordinal position. Although patients were slower than controls **(middle)**, the rate of retrieval was similar to that of controls **(right)**. (Adapted from McDougle et al., 2022).

The importance of the cerebellum in continuous spatial processing is consistent with the evidence from many other tasks and measures. In terms of perceptual tasks, individuals with CA have difficulty judging the velocity of a moving object (87). Similarly, individuals with CA struggle to predict when and where an object will reappear after passing behind an occluder (76) (also see: (77)). Preliminary work from our lab indicates that these individuals also have difficulty on intuitive physics problems that require using spatiotemporal information to predict the trajectory of an object being released from a pendulum (88). Here, too, we observe a parametric effect where the impairment becomes more pronounced as the extrapolated distance is increased.

The use of spatial representations and transformations on these representations is fundamental to many of our cognitive abilities such as mathematics, linguistic relationships, and metaphorical thinking (89–91). To test whether the cerebellum supports continuous spatial representational transformations in more abstract domains, we compared CA individuals and controls on simple addition and multiplication problems (Figure 5). Single-digit addition is thought to involve a mental traversal along an internal, spatialized number line (92–95). Thus, verifying that $4 + 3 = 7$ takes less time than verifying that $8 + 6 = 14$ because the mental distance traveled is shorter (96, 97). In contrast, solving single-digit multiplication problems is assumed to rely on a discrete look-up table, with faster verification for problems like $4 \times 3 = 12$ than $8 \times 6 = 48$ due to stronger memory traces accumulated over a lifetime of practice.

Based on this computational distinction, we predicted that individuals with CA would show an elevated RT slope for the addition problems but not for the multiplication problems. The results were consistent with this prediction: As the problem size increased, individuals with CA showed a steeper RT slope than matched controls on the addition problems but not on the multiplication problems (85). Unlike the earlier comparison between mental rotation and memory recall—where visual stimuli and task demands differed substantially—the addition and multiplication tasks used visually matched operands and identical verification (True vs False) responses across conditions. We interpret this dissociation as arising because only the addition task requires an internal, continuous transformation—a process dependent on cerebellar computation.

In a follow-up study, we used problems with either two or three addends (98). Again, the CA group showed an elevated slope when plotting RT as a function of problem size. However, the CA group showed a similar RT increase as controls when moving from two- to three-addend problems. We interpret this as a case where discrete and continuous operations are combined within the same task: The RT increase with problem size reflects a continuous transformation along a mental number line, whereas the additional RT for three-addend problems reflects discrete retrieval of the third item after summing the first two. Interestingly, individuals with Parkinson's disease show the opposite pattern: A greater increase in RT compared to controls (and CA) when going from two to three addends, but a similar slope as controls as a function of the problem size. This double dissociation reinforces the cerebellum's specificity for predictions involving continuous, not discrete, internal transformations.

This section has emphasized the Continuity constraint. Mental rotation and simple addition—tasks in which efficient processing depends on cerebellar integrity—involve continuous transformation of a mental representation. In contrast, spatial working memory and multiplication—tasks unaffected by cerebellar dysfunction—rely on retrieving discrete entities from working or long-term memory. Nonetheless, the Prediction and Timescale constraints are also operative here. Mental rotation is inherently predictive—the current state is internally transformed onto a future state, and within the context of mental rotation, the default prediction is that the stimulus will be in its normal orientation. Similarly, movement along a number line predicts the transformed value of an initial number (e.g., "Does adding 3 to 5 yield 8?"). These operations also unfold over a limited timescale (e.g., mental rotation rates of around 2 ms/deg for controls and 3 ms/deg for the CA group). Altogether, these findings again strongly support CoRT, underscoring the cerebellum's specialized role in generating continuous predictions over short intervals.

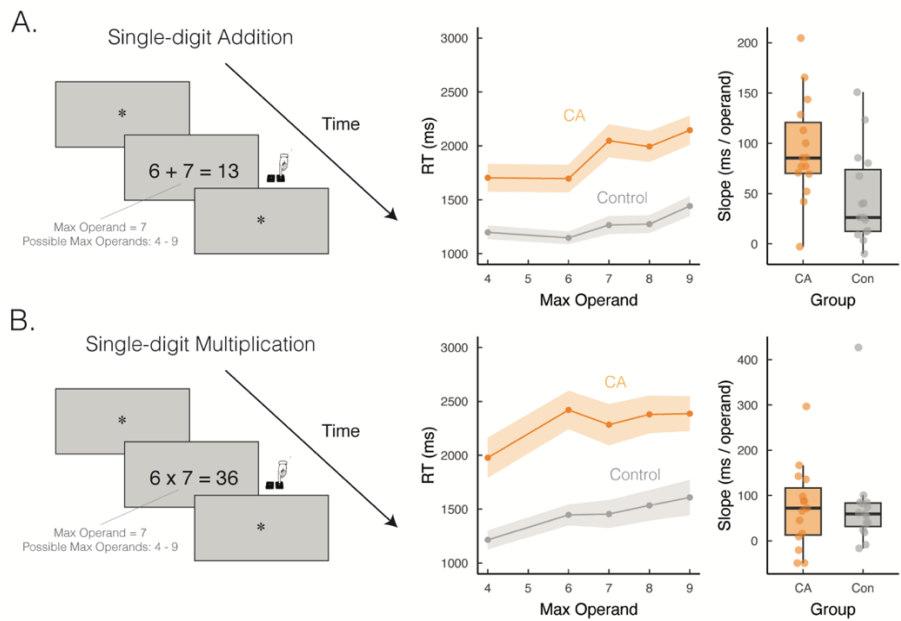

**Figure 5. Cerebellar involvement in continuous numerical cognition.** Participants made speeded true/false judgments on arithmetic equations **(left)** involving single-digit **(A)** addition or **(B)** multiplication. Simple addition is assumed to rely on reference to a continuous, spatialized number line; simple multiplication is assumed to rely on retrieval from a discretized look-up table. Patients with cerebellar ataxia were slower on both tasks **(middle)** but only showed an increase in slope on the addition task **(right)** (adapted from McDougle et al., 2022).

## IV. Extending CoRT to Other Cognitive Domains

Moving forward, we outline how the three CoRT constraints can guide future research across other cognitive domains. Specifically, we focus on language, social cognition, and cognitive control—areas extensively studied in the cerebellar literature, yet still lacking a clear account of the cerebellum's functional role.

Interest in the role of the cerebellum in language traces back to the earliest neuroimaging studies. A seminal PET study looking at the stages of semantic processing showed an increase in cerebellar activation in right Crus I and II when participants were asked to covertly generate a semantic associate to a target word compared to when they silently read the target word (99, 100). This activation was puzzling given that the sensorimotor requirements were equated between conditions—silently producing a single word. When subsequent studies showed cerebellar activation in a broad range of language tasks, early functional hypotheses proposed that the cerebellum was a node in the covert speech network supporting the maintenance of information in verbal working memory (101, 102).

However, the results from experiments designed to test this hypothesis failed to support this motor-centric account of the cerebellum's role in language, inspiring researchers to consider alternative hypotheses. One prominent hypothesis centered on the idea that the cerebellum operated as a generalized forward model, using contextual cues—whether read or heard—to anticipate upcoming linguistic inputs (29, 31). Consistent with this view, BOLD activity in language-associated cerebellar regions increases during a sentence verification task when the final word is predictable (Figure 6A-B). Moreover, violations of semantic predictions evoke a robust cerebellar response, a response interpreted as an error signal (103, 104).

Inferences based on the imaging work need to be tempered when considering neuropsychological data. On a sentence verification task, individuals with CA showed the same reaction time advantage as controls when the final word was highly predictive versus weakly predictive (Figure 6C) (105). Inspired by CoRT, we also manipulated the vividness of spatial imagery of the sentences, reasoning that the cerebellum may preferentially support linguistic predictions when language invites internal spatial imagery and simulation. Here again, the RT slope for the CA group was similar to that of controls when comparing high and low imagery sentences.

This latter result could certainly be viewed as problematic for CoRT, at least in terms of its relevance for understanding predictive functions of the cerebellum in the language domain. Alternatively—and admittedly post hoc—the advent of large language models suggests that massive discrete lookup tables may suffice to generate linguistic predictions, obviating the need for internal transformations in semantic space (106). Retrieval operations, such as those central to these large language models, appear to be preserved across multiple task domains in CA.

To test this idea, future work could assess whether cerebellar patients show selective impairments on semantic tasks requiring internal transformation. For example, simple retrieval tasks (e.g., completing "salt and ___") can be solved by accessing a stored association. In contrast, analogy tasks (e.g., "king is to queen as man is to ___") go beyond retrieving definitions, requiring the transformation of one concept into another along a graded semantic dimension (107, 108). Comparing these conditions in cerebellar patients could clarify whether the cerebellum supports language prediction specifically when it relies on continuous internal transformations.

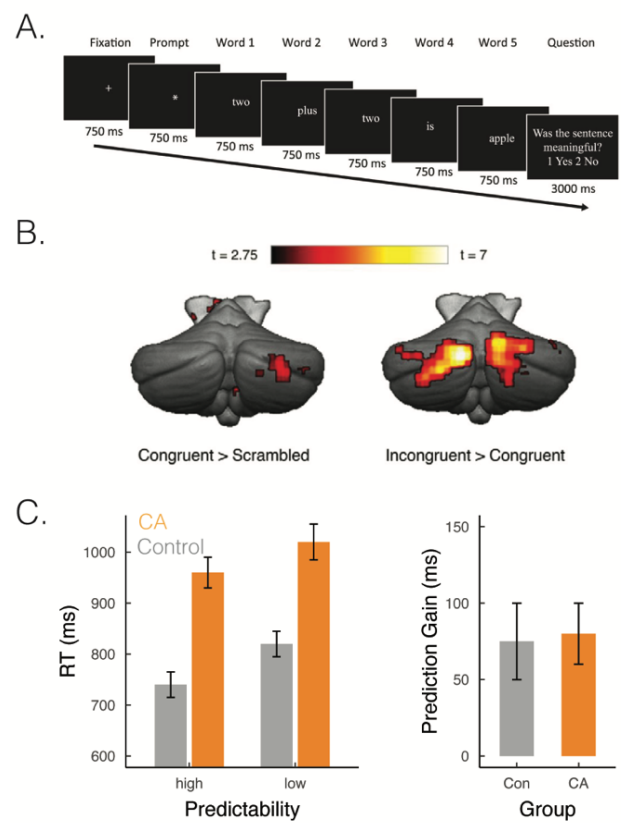

**Figure 6. Does the cerebellum operate as a forward model in the language domain? A)** Participants judged whether sentences were meaningful or not. Congruent trials ended with contextually appropriate words (e.g., The pizza was too hot to eat.), incongruent trials contained semantically anomalous endings (as shown), and scrambled trials presented the same words in a random order. **B)** Cerebellar BOLD was higher in right Crus I when the final word was predictable compared to the scrambled condition and higher bilaterally when the final word violated a prediction (adapted from Moberget, et al. 2014). **C)** Performance on a similar task by patients with cerebellar ataxia and controls, using sentences in which predictability of the final word on congruent trials was either high or low. Although patients were slower, they showed same benefit (prediction gain) on trials with high predictability (adapted from King et al., 2024). While the fMRI data would point to a role of the cerebellum in generating semantic predictions, the behavioral data suggest that semantic prediction is intact in the patients.

Social cognition offers another domain ripe for extending CoRT. Neuroimaging studies consistently show cerebellar activation—particularly in the posterior cerebellum (right Crus I and II)—during tasks involving mentalizing and perspective-taking (32, 109). These data have also been interpreted within the framework of a generalized forward model, with the cerebellum essential for simulating an event sequence that would allow one to infer the mental state and future behavior of other individuals. Moreover, it has been argued that the cerebellar contribution is especially critical in social contexts; for example, cerebellar activation is especially pronounced when the picture sequence requires assessing the perspective of another individual (i.e., theory of mind task) (3, 109, 110).

The neuropsychological data provide further evidence of a role for the cerebellum in social cognition. Changes in social and affective behavior, primarily measured with subjective reports provided by the patients and their caregivers, are features of CCAS. Experimentally, individuals with cerebellar degeneration show impairments on the picture sequencing task when the story depicts a scenario that requires simulating others' dynamic mental states (111). Less direct evidence comes from the psychiatry literature where cerebellar dysfunction has been hypothesized to be one of the causal mechanisms underlying the social deficits observed in autism (112) and schizophrenia (113). Animal models provide converging evidence showing that cerebellar damage can impair social behavior such as a preference for conspecifics relative to inanimate objects (10).

However, as discussed previously, how the cerebellum supports social cognition through its predictive capacity remains *underspecified*: Does the cerebellum play a specific role in continuous sequential processing, something common in social scenarios? Or is it the social nature of the tasks that engages the cerebellum? We believe the core constraints of Continuity, Prediction, and Timescale offer a path forward in addressing these questions. For example, one could contrast tasks that require continuous transformations such as tracking evolving emotional states (114), with those involving static social knowledge such as identifying social roles or recalling biographical facts. Or a comparison could be made between conditions in which continuous sequential processing is assessed in social and non-social contexts, or in which social scenarios unfold over different time scales. The CoRT notion would predict dissociations in the patients' performance, with deficits limited to sequential processing tasks that require continuous transformations over a limited timescale, social or otherwise. Such a finding would not discount the importance of the cerebellum in social cognition. It may well be that these types of computations are especially essential for maintaining cohesive social behavior.

Cognitive control is another promising domain for exploring the utility of CoRT. Tasks requiring control processes such as flexible reasoning, multitasking, and performance monitoring activate the cerebellum; indeed, cognitive control is featured as one of the core cerebellar networks in a recent functional map (13). Impairments in cognitive control are at the heart of the cognitive deficits measured with the CCAS, with the clinical picture sometimes described as a mild form of frontal lobe syndrome. Yet, as reviewed above, patients with CA search through spatial working memory at the same rate as controls, a result that is at odds with a general cognitive control deficit.

With our collaborators, we recently proposed an alternative approach to analyzing cerebellar fMRI activity: using a functional map of cortico-cerebellar connectivity to identify conditions where the cerebellar BOLD response *exceeds* that predicted from its cortical inputs (115, 116). Consistent with the patient work, changes in cerebellar activation as the demands on working memory retrieval increased were fully accounted for by the changes in cortical input. In contrast, the cerebellar activation increased more than predicted as the demands on encoding into working memory increased.

Future studies could be designed to test if Prediction, Time-scale, and/or Continuity are more relevant for the mental operations engaged during encoding compared to retrieval. For example, one could contrast an encoding task that requires short time-scale predictions such as phonological sequencing, with an encoding task that requires long time-scale integration such as narrative comprehension. The continuity constraint could be tested by comparing encoding when the memory set consists of shapes that morph smoothly into one another versus when the shapes are presented as discrete and categorical. The CoRT notion would predict dissociations in the patients' performance, with deficits limited to encoding tasks that require continuous transformations over a limited timescale.

In summary, we propose that CoRT offers a useful starting point for generating hypotheses about cerebellar involvement in non-motor domains, especially where there are discrepancies between findings from different methodologies.

## V. Toward a Deeper Understanding of the Cerebellum

We close this perspective by briefly commenting on a few meta-issues that require special attention as the field continues to explore the role of the cerebellum in cognition. These issues are not specifically related to the CoRT framework. Rather, we highlight general insights that emerge from the cerebellum and cognition literature which may prove useful in developing and evaluating mechanistic hypotheses.

**Need for Converging Evidence**. As with all scientific enterprises, the quest to understand the functional domain of the cerebellum entails evidence drawn from multiple methodologies. Interestingly, whereas behavioral neurology laid out foundational ideas for much of cortical function, the seeds of the cerebellum and cognition revolution stemmed from neuroanatomy and neuroimaging. Indeed, serious consideration that the functional domain extended beyond the motor domain took hold with the repeated observations in the neuroimaging literature of activation patterns that could not be explained in terms of motoric demands (99, 117, 118). Prior to the imaging work, there was almost no reference in the clinical literature of cognitive or affective changes associated with cerebellar disorders.

This last point underscores an important point. The changes in cognition observed in individuals with late-onset cerebellar degeneration are typically quite subtle, especially in subtypes with minimal extracerebellar involvement. We don't see a striking loss of function—a loss of the ability to speak, neglect of a hemifield, or inability to recall what happened a few minutes ago. This absence of impairment is even more striking in individuals who experience a cerebellar stroke or undergo surgery that entails removal of cerebellar tissue (119). Once past the acute phase, these individuals frequently score within normal bounds on most neuropsychological assessments (120–122).

More generally, there is frequently a misalignment between the neuroimaging and neuropsychological data. We've noted how the cerebellum shows pronounced activation when semantic predictions are violated, yet patients with CA appear to show similar sensitivity to semantic predictability as controls. This discrepancy extends to other domains, including spatial attention and reinforcement learning. In the latter, neuroimaging and neurophysiology reveal cerebellar signatures of reward prediction errors (9, 123–126), yet patients with CA tend to perform normally on standard reinforcement learning tasks (127–129) (But see: (130)).

We can think of at least three explanations for this discrepancy. First, the mismatch may underscore the need for more sophisticated psychophysical tasks and behavioral analyses. Neuropsychological batteries are relatively crude instruments, designed to efficiently provide a profile of impairment and spared function. They lack the sensitivity to detect the types of problems that may arise when a damaged neural system provides the computations that optimize a particular cognitive operation. For example, in our mental rotation study, the patients were slower at mentally rotating the images but, with the extra time, still came up with the right answer. The same holds for behavioral assays in animal studies. While the social cognition tasks used with rodents often quantify social preference as time spent with other animals versus objects, more fine-grained behavioural dissections may be needed to identify the specific constraints—such as continuity, prediction, or timescale—that engage the cerebellum during social cognition.

Relevant to this last point is another observation that demands careful consideration in future neuropsychological experiments. In evaluating the relevance of CoRT-related constraints in the spatial domain, we focused on the slope measure, taking this as indicative of the fluidity or rate at which individuals manipulate a mental representation (e.g., mental rotation). The absence of a slope difference was critical to our claim that CA patients were not impaired in the rate at which they performed non-continuous transformations (e.g., iterative retrieval from working memory).

But in all conditions, there was a main effect with the CA group showing a marked increase in RT compared to the control group. This increase, on the order of 300-500 ms even in the simplest mental rotation or working memory conditions is well beyond what one might expect from movement problems related to their ataxia. For example, on simple detection tasks, individuals with CA show only a 50 ms increase in RT. Thus, it appears that, even when a particular cognitive operation appears spared (e.g., rate of search through a discrete look-up table), CA is associated with a general increase in processing time as the cognitive demands increase. This cost, should it be generic across all task domains, would be hard to reconcile within a model that limits cerebellar processing to operations that conform to the prediction, continuity, and timescale constraints. An important challenge for future work is to develop mechanistic hypotheses that can account for both general and specific forms of impairment.

Second, caution is warranted when drawing inferences about cerebellar function from neuroimaging data. While this caveat applies broadly, it is especially critical for cerebellar studies as the cerebellar BOLD response primarily reflects input from the pontine nuclei. As such, these data tell us about the information that the cerebellum has access to but shed little light on the computations performed by the Purkinje cells and deep cerebellar nuclei. We noted above one alternative approach for cerebellar neuroimaging, testing whether the cerebellar activity is greater than expected given its input (Shahshahani et al. 2024). The underlying assumption here is that the input might be gated to amplify information that will be essential for cerebellar-specific computations.

Third, the patients may show minimal or modest impairment because of reorganization and compensation. By this view, either spared tissue in the cerebellum or extracerebellar structures have come to compensate for the loss function. This is a general issue of concern in neuropsychological research. While compensatory hypotheses should certainly be considered, it seems less likely to be relevant when considering disorders associated with slow, degenerative processes, especially when the degeneration is widespread across the cerebellum. We are unaware of compelling evidence for such compensation in degenerative diseases (e.g., individuals with Alzheimer's disease do not generate alternative memory pathways).

Experimentally, an important way to address the compensatory issue is to use methods in which cerebellar function is transiently disrupted. Optogenetics provides a powerful tool for work with non-human models (10, 41), although the behavioral tasks tend to not be very analytic. In human experimental work, there has been a recent avalanche of studies using non-invasive stimulation methods in humans (131, 132). Much of that work has been directed at the descriptive level, using these interventions as another way to demonstrate that cerebellar perturbations disrupt performance in particular task domains. Combining these methods with sophisticated behavioral assays is certainly a promising direction for future research.

The compensation issue points to another puzzle for researchers interested in cerebellar function. Insult to the cerebellum early in life appears to produce more marked impairments in cognition and affect compared to when the pathology arises during adulthood (as with most degenerative disorders) (119, 133). Moreover, early insult is associated with cognitive and behavioral issues that can last the individual's lifetime; for example, early cerebellar damage is the second-highest risk factor for autism (134). This pattern is especially intriguing given that cortical lesions present the opposite picture: For the cortex, the prognosis for recovery is better when the damage occurs earlier in life compared to later in life. For example, children who undergo hemispherectomy for severe epilepsy—removal of half the brain—often show remarkable plasticity, with cognitive abilities only slightly below those of their peers (135). In contrast, similar cortical lesions later in life result in irreversible deficits (136).

The observation that the kinds of severe cognitive and emotional impairments observed with prenatal or childhood cerebellar insult are not seen with adults raises intriguing questions about inferring normal cerebellar function. It may be that the cerebellum plays a critical role early in development to optimize processing in extracerebellar pathways. However, once that processing system is established, the cerebellar contribution may no longer be needed. Under this scenario, studying cognitive and affective changes associated with childhood insult would be important to understand how the cerebellum contributes to development but may have limited value for inferring function of the adult cerebellum.

**Beyond the Universal Cerebellar Transform and Functional Specialization**. In the search for a unifying theory, many researchers—including ourselves—have been guided by the idea of a universal cerebellar transform (UCT): the notion that the cerebellum implements a specialized computational algorithm that is applied broadly across motor and cognitive domains. This idea was motivated by the strikingly uniform anatomy and physiology of the cerebellum, coupled with its extensive connectivity with the cerebral cortex— features that suggest a domain-general, algorithmic function (Marr, 1969; Albus, 1971; Eccles, 1973; Strick, Dum, & Fiez, 2009; King et al., 2023). Although varying in their level of specificity, candidate UCTs have focused on computations related to error correction (26), timing (137), homeostatic maintenance (2), and forward modeling (23).

Empirically, recent neurophysiological findings suggest that the picture is more nuanced. While the cerebellum certainly lacks the cellular diversity of the cerebral cortex, there is considerable anatomical heterogeneity in terms of microcircuitry with distinct microzones (138, 139). Compounding this, even within the motor domain, signals represented in the cerebellum are more diverse than previously assumed. For example, in addition to providing an error signal, complex spikes encode higher-order variables such as reward and goal states—features that would be difficult to capture within a narrow computational framework (Orban de Xivry & Diedrichsen, 2024).

It is also possible that the very theoretical construct of UCT is misguided. Instead of searching for a singular computation shared across task domains, it may be more fruitful to examine how diverse cerebellar circuits carry out distinct computations. Alternatively, it may be that the cerebellum supports cortical processing through its sheer number of neurons that can facilitate pattern expansion and recognition (140). By this view, the cerebellum need not be considered to represent something per se. Rather, in the way that oil is essential to the smooth operation of an engine, the cerebellum might ensure that the mind operates fluidly and efficiently, what might be described as mental coordination.

Here, we advance an integrated perspective. Rather than positing a single unifying algorithm or treating cerebellar computation as bespoke to each domain, we highlight a set of constraints that consistently characterize cerebellar processing. Like mathematical postulates, these constraints—together with those yet to be identified—can be precisely defined, empirically tested, and used to evaluate consistencies and discrepancies across functional domains, moving us closer to a comprehensive understanding of the cerebellum's role in cognition and action.

# References

1. H. C. Leiner, A. L. Leiner, R. S. Dow, Does the cerebellum contribute to mental skills? *Behav. Neurosci.* **100**, 443–454 (1986).

2. J. D. Schmahmann, X. Guell, C. J. Stoodley, M. A. Halko, The Theory and Neuroscience of Cerebellar Cognition. *Annu. Rev. Neurosci.* **42**, 337–364 (2019).

3. A. A. Sokolov, R. C. Miall, R. B. Ivry, The Cerebellum: Adaptive Prediction for Movement and Cognition. *Trends Cogn. Sci.* **21**, 313–332 (2017).

4. H. Jacobi, J. Faber, D. Timmann, T. Klockgether, Update cerebellum and cognition. *J. Neurol.* **268**, 3921–3925 (2021).

5. M. King, C. R. Hernandez-Castillo, R. A. Poldrack, R. B. Ivry, J. Diedrichsen, Functional boundaries in the human cerebellum revealed by a multi-domain task battery. *Nat. Neurosci.* **22**, 1371–1378 (2019).

6. F. Hoche, X. Guell, M. G. Vangel, J. C. Sherman, J. D. Schmahmann, The cerebellar cognitive affective/Schmahmann syndrome scale. *Brain* **141**, 248–270 (2018).

7. J. D. Schmahmann, J. C. Sherman, The cerebellar cognitive affective syndrome. *Brain* **121 ( Pt 4)**, 561–579 (1998).

8. S. Rudolph, *et al.*, Cognitive-affective functions of the cerebellum. *J. Neurosci.* **43**, 7554–7564 (2023).

9. D. Kostadinov, M. Häusser, Reward signals in the cerebellum: Origins, targets, and functional implications. *Neuron* **110**, 1290–1303 (2022).

10. I. Carta, C. H. Chen, A. L. Schott, S. Dorizan, K. Khodakhah, Cerebellar modulation of the reward circuitry and social behavior. *Science* **363** (2019).

11. C. Hull, Prediction signals in the cerebellum: beyond supervised motor learning. *Elife* **9** (2020).

12. J. X. O'Reilly, C. F. Beckmann, V. Tomassini, N. Ramnani, H. Johansen-Berg, Distinct and overlapping functional zones in the cerebellum defined by resting state functional connectivity. *Cereb. Cortex* **20**, 953–965 (2010).

13. C. Nettekoven, *et al.*, A hierarchical atlas of the human cerebellum for functional precision mapping. *Nat. Commun.* **15**, 8376 (2024).

14. I. Sugihara, Y. Shinoda, Molecular, topographic, and functional organization of the cerebellar nuclei: analysis by three-dimensional mapping of the olivonuclear projection and aldolase C labeling. *J. Neurosci.* **27**, 9696–9710 (2007).

15. J. Voogd, J. Pardoe, T. J. H. Ruigrok, R. Apps, The distribution of climbing and mossy fiber collateral branches from the copula pyramidis and the paramedian lobule: congruence of climbing fiber cortical zones and the pattern of zebrin banding within the rat cerebellum. *J. Neurosci.* **23**, 4645–4656 (2003).

16. N. L. Cerminara, E. J. Lang, R. V. Sillitoe, R. Apps, Redefining the cerebellar cortex as an assembly of non-uniform Purkinje cell microcircuits. *Nat. Rev. Neurosci.* **16**, 79–93 (2015).

17. D. Marr, A theory of cerebellar cortex. *J. Physiol.* **202**, 437–470 (1969).

18. J. B. Heald, M. Lengyel, D. M. Wolpert, Contextual inference in learning and memory. *Trends Cogn. Sci.* **27**, 43–64 (2023).

19. A. Collins, E. Koechlin, Reasoning, learning, and creativity: frontal lobe function and human decision-making. *PLoS Biol.* **10**, e1001293 (2012).

20. J. F. Medina, K. S. Garcia, W. L. Nores, N. M. Taylor, M. D. Mauk, Timing mechanisms in the cerebellum: testing predictions of a large-scale computer simulation. *J. Neurosci.* **20**, 5516–5525 (2000).

21. M. Ito, M. Kano, Long-lasting depression of parallel fiber-Purkinje cell transmission induced by conjunctive stimulation of parallel fibers and climbing fibers in the cerebellar cortex. *Neurosci. Lett.* **33**, 253–258 (1982).

22. J. S. Albus, A theory of cerebellar function. *Math. Biosci.* **10**, 25–61 (1971).

23. D. M. Wolpert, R. C. Miall, M. Kawato, Internal models in the cerebellum. *Trends Cogn. Sci.* **2**, 338–347 (1998).

24. N. H. Barmack, V. Yakhnitsa, Functions of interneurons in mouse cerebellum. *J. Neurosci.* **28**, 1140–1152 (2008).

25. M. A. Gaffield, J. M. Christie, Movement rate is encoded and influenced by widespread, coherent activity of cerebellar molecular layer interneurons. *J. Neurosci.* **37**, 4751–4765 (2017).

26. J. L. Raymond, S. G. Lisberger, M. D. Mauk, The cerebellum: a neuronal learning machine? *Science* **272**, 1126–1131 (1996).

27. M. Ito, Movement and thought: identical control mechanisms by the cerebellum. *Trends in Neurosciences* [Preprint] (1993). Available at: http://dx.doi.org/10.1016/0166-2236(93)90073-u.

28. M. Haruno, D. M. Wolpert, M. Kawato, Mosaic model for sensorimotor learning and control. *Neural Comput.* **13**, 2201–2220 (2001).

29. A. LeBel, A. M. D'Mello, A seat at the (language) table: incorporating the cerebellum into frameworks for language processing. *Current Opinion in Behavioral Sciences* **53**, 101310 (2023).

30. C. Casto, *et al.*, The cerebellar components of the human language network. *bioRxiv* 2025.04.14.645351 (2025).

31. T. Moberget, R. B. Ivry, Cerebellar contributions to motor control and language comprehension: searching for common computational principles. *Ann. N. Y. Acad. Sci.* **1369**, 154–171 (2016).

32. F. Van Overwalle, *et al.*, Consensus paper: Cerebellum and social cognition. *Cerebellum* **19**, 833–868 (2020).

33. J. S. Tsay, *et al.*, Fundamental processes in sensorimotor learning: Reasoning, refinement, and retrieval. *Elife* **13** (2024).

34. J. Krakauer, A. M. Hadjiosif, J. Xu, A. L. Wong, A. M. Haith, Motor Learning. *Compr. Physiol.* **9**, 613–663 (2019).

35. K. Friston, S. Kiebel, Predictive coding under the free-energy principle. *Philos. Trans. R. Soc. Lond. B Biol. Sci.* **364**, 1211–1221 (2009).

36. S. M. Morton, A. J. Bastian, Cerebellar contributions to locomotor adaptations during splitbelt treadmill walking. *J. Neurosci.* **26**, 9107–9116 (2006).

37. D. A. Nowak, J. Hermsdörfer, K. Rost, D. Timmann, H. Topka, Predictive and reactive finger force control during catching in cerebellar degeneration. *Cerebellum* **3**, 227–235 (2004).

38. G. Holmes, The symptoms of acute cerebellar injuries due to gunshot injuries. *Brain* **40**, 461–535 (1917).

39. C. M. Gomez, *et al.*, Spinocerebellar ataxia type 6: gaze-evoked and vertical nystagmus, Purkinje cell degeneration, and variable age of onset. *Ann. Neurol.* **42**, 933–950 (1997).

40. R. C. Miall, L. O. D. Christensen, O. Cain, J. Stanley, Disruption of state estimation in the human lateral cerebellum. *PLoS Biol.* **5**, e316 (2007).

41. M. I. Becker, A. L. Person, Cerebellar control of reach kinematics for endpoint precision. *Neuron* **103**, 335-348.e5 (2019).

42. N. Sinha, *et al.*, Disentangling acute motor deficits and adaptive responses evoked by the loss of cerebellar output. *Elife* **14** (2025).

43. A. J. Bastian, Learning to predict the future: the cerebellum adapts feedforward movement control. *Curr. Opin. Neurobiol.* **16**, 645–649 (2006).

44. J. S. Tsay, H. Kim, A. M. Haith, R. B. Ivry, Understanding implicit sensorimotor adaptation as a process of proprioceptive re-alignment. *Elife* **11** (2022).

45. R. Shadmehr, M. A. Smith, J. Krakauer, Error correction, sensory prediction, and adaptation in motor control. *Annu. Rev. Neurosci.* **33**, 89–108 (2010).

46. P. Mazzoni, J. W. Krakauer, An implicit plan overrides an explicit strategy during visuomotor adaptation. *J. Neurosci.* **26**, 3642–3645 (2006).

47. H. Golla, *et al.*, Reduced saccadic resilience and impaired saccadic adaptation due to cerebellar disease. *Eur. J. Neurosci.* **27**, 132–144 (2008).

48. K. Rabe, *et al.*, Adaptation to visuomotor rotation and force field perturbation is correlated to different brain areas in patients with cerebellar degeneration. *J. Neurophysiol.* **101**, 1961–1971 (2009).

49. M. Hallett, B. T. Shahani, R. R. Young, EMG analysis of patients with cerebellar deficits. *J. Neurol. Neurosurg. Psychiatry* **38**, 1163–1169 (1975).

50. A. S. Therrien, D. M. Wolpert, A. J. Bastian, Effective reinforcement learning following cerebellar damage requires a balance between exploration and motor noise. *Brain* **139**, 101–114 (2016).

51. J. S. Tsay, L. Schuck, R. B. Ivry, Cerebellar degeneration impairs strategy discovery but not strategy recall. *Cerebellum* (2022). https://doi.org/10.1007/s12311-022-01500-6.

52. A. L. Wong, C. L. Marvel, J. A. Taylor, J. W. Krakauer, Can patients with cerebellar disease switch learning mechanisms to reduce their adaptation deficits? *Brain* (2019). https://doi.org/10.1093/brain/awy334.

53. B. Parrell, H. E. Kim, A. Breska, A. Saxena, R. B. Ivry, Differential effects of cerebellar degeneration on feedforward versus feedback control across speech and reaching movements. *J. Neurosci.* (2021). https://doi.org/10.1523/JNEUROSCI.0739-21.2021.

54. F. B. Horak, H. C. Diener, Cerebellar control of postural scaling and central set in stance. *J. Neurophysiol.* **72**, 479–493 (1994).

55. A. M. Zimmet, D. Cao, A. J. Bastian, N. J. Cowan, Cerebellar patients have intact feedback control that can be leveraged to improve reaching. *Elife* **9** (2020).

56. N. H. Bhanpuri, A. M. Okamura, A. J. Bastian, Predictive modeling by the cerebellum improves proprioception. *J. Neurosci.* **33**, 14301–14306 (2013).

57. K. E. Cullen, Internal models of self-motion: neural computations by the vestibular cerebellum. *Trends Neurosci.* **46**, 986–1002 (2023).

58. R. Morehead, J. A. Taylor, D. E. Parvin, R. B. Ivry, Characteristics of Implicit Sensorimotor Adaptation Revealed by Task-irrelevant Clamped Feedback. *J. Cogn. Neurosci.* **29**, 1061–1074 (2017).

59. Y.-W. Tseng, J. Diedrichsen, J. W. Krakauer, R. Shadmehr, A. J. Bastian, Sensory prediction errors drive cerebellum-dependent adaptation of reaching. *J. Neurophysiol.* **98**, 54–62 (2007).

60. A. Breska, R. B. Ivry, Taxonomies of timing: Where does the cerebellum fit in? *Curr. Opin. Behav. Sci.* **8**, 282–288 (2016).

61. R. F. Thompson, The neurobiology of learning and memory. *Science* **233**, 941–947 (1986).

62. D. S. Woodruff-Pak, M. Papka, R. B. Ivry, Cerebellar involvement in eyeblink classical conditioning in humans. *Neuropsychology* **10**, 443 (1996).

63. M. Gerwig, *et al.*, Trace eyeblink conditioning in patients with cerebellar degeneration: comparison of short and long trace intervals. *Exp. Brain Res.* **187**, 85–96 (2008).

64. H. Topka, J. Valls-Solé, S. G. Massaquoi, M. Hallett, Deficit in classical conditioning in patients with cerebellar degeneration. *Brain* **116**, 961–969 (1993).

65. I. Daum, *et al.*, Classical conditioning after cerebellar lesions in humans. *Behav. Neurosci.* **107**, 748–756 (1993).

66. D. Timmann, *et al.*, The human cerebellum contributes to motor, emotional and cognitive associative learning. A review. *Cortex* **46**, 845–857 (2010).

67. D. G. Lavond, J. S. Lincoln, D. A. McCormick, R. F. Thompson, Effect of bilateral lesions of the dentate and interpositus cerebellar nuclei on conditioning of heart-rate and nictitating membrane/eyelid responses in the rabbit. *Brain Res.* **305**, 323–330 (1984).

68. J. F. Medina, M. D. Mauk, Computer simulation of cerebellar information processing. *Nat. Neurosci.* **3 Suppl**, 1205–1211 (2000).

69. K. P. Nguyen, A. L. Person, Cerebellar circuit computations for predictive motor control. *Nat. Rev. Neurosci.* (2025). https://doi.org/10.1038/s41583-025-00936-z.

70. D. Narain, E. D. Remington, C. I. D. Zeeuw, M. Jazayeri, A cerebellar mechanism for learning prior distributions of time intervals. *Nat. Commun.* **9** (2018).

71. R. B. Ivry, S. W. Keele, Timing functions of the cerebellum. *J. Cogn. Neurosci.* **1**, 136–152 (1989).

72. G. Koch, *et al.*, Repetitive TMS of cerebellum interferes with millisecond time processing. *Exp. Brain Res.* **179**, 291–299 (2007).

73. K.-H. Lee, *et al.*, The role of the cerebellum in subsecond time perception: evidence from repetitive transcranial magnetic stimulation. *J. Cogn. Neurosci.* **19**, 147–157 (2007).

74. H. Ackermann, S. Gräber, I. Hertrich, I. Daum, Categorical speech perception in cerebellar disorders. *Brain Lang.* **60**, 323–331 (1997).

75. R. B. Ivry, S. W. Keele, H. C. Diener, Dissociation of the lateral and medial cerebellum in movement timing and movement execution. *Exp. Brain Res.* **73**, 167–180 (1988).

76. M. J. Roth, M. Synofzik, A. Lindner, The cerebellum optimizes perceptual predictions about external sensory events. *Curr. Biol.* **23**, 930–935 (2013).

77. J. X. O'Reilly, M. M. Mesulam, A. C. Nobre, The cerebellum predicts the timing of perceptual events. *J. Neurosci.* **28**, 2252–2260 (2008).

78. M. Grube, F. E. Cooper, P. F. Chinnery, T. D. Griffiths, Dissociation of duration-based and beat-based auditory timing in cerebellar degeneration. *Proc. Natl. Acad. Sci. U. S. A.* **107**, 11597–11601 (2010).

79. A. Breska, R. B. Ivry, The human cerebellum is essential for modulating perceptual sensitivity based on temporal expectations. *Elife* **10** (2021).

80. A. Breska, R. B. Ivry, Double dissociation of single-interval and rhythmic temporal prediction in cerebellar degeneration and Parkinson's disease. *Proceedings of the National Academy of Sciences* **115**, 12283–12288 (2018).

81. M. J. Tarr, S. Pinker, Mental rotation and orientation-dependence in shape recognition. *Cogn. Psychol.* **21**, 233–282 (1989).

82. R. N. Shepard, J. Metzler, Mental rotation of three-dimensional objects. *Science* **171**, 701–703 (1971).

83. J. M. Zacks, Neuroimaging studies of mental rotation: a meta-analysis and review. *J. Cogn. Neurosci.* **20**, 1–19 (2008).

84. A. P. Georgopoulos, J. T. Lurito, M. Petrides, A. B. Schwartz, J. T. Massey, Mental rotation of the neuronal population vector. *Science* **243**, 234–236 (1989).

85. S. D. McDougle, *et al.*, Continuous manipulation of mental representations is compromised in cerebellar degeneration. *Brain* (2022). https://doi.org/10.1093/brain/awac072.

86. S. Sternberg, High-speed scanning in human memory. *Science* **153**, 652–654 (1966).

87. R. B. Ivry, H. C. Diener, Impaired velocity perception in patients with lesions of the cerebellum. *J. Cogn. Neurosci.* **3**, 355–366 (1991).

88. T. Wang, M. Xia, R. Ivry, Patients with Cerebellar Ataxia exhibit deficiencies in simulating physical dynamics. (2024).

89. G. Lakoff, M. Johnson, The metaphorical structure of the human conceptual system. *Cogn. Sci.* **4**, 195–208 (1980).

90. D. Casasanto, *Space for thinking,"* in *Language, Cognition, and Space: State of the Art and New Directions*, V. Evans, P. Chilton, Eds. (Equinox Publishing, 2010).

91. B. Pitt, A. Carstensen, I. Boni, S. T. Piantadosi, E. Gibson, Different reference frames on different axes: Space and language in indigenous Amazonians. *Sci. Adv.* **8**, eabp9814 (2022).

92. S. Göbel, V. Walsh, M. F. Rushworth, The mental number line and the human angular gyrus. *Neuroimage* **14**, 1278–1289 (2001).

93. R. Mathieu, A. Gourjon, A. Couderc, C. Thevenot, J. Prado, Running the number line: Rapid shifts of attention in single-digit arithmetic. *Cognition* **146**, 229–239 (2016).

94. K. Miller, M. Perlmutter, D. Keating, Cognitive arithmetic: comparison of operations. *J. Exp. Psychol. Learn. Mem. Cogn.* **10**, 46–60 (1984).

95. M. Zorzi, K. Priftis, C. Umiltà, Brain damage: neglect disrupts the mental number line. *Nature* **417**, 138–139 (2002).

96. K. Mccrink, S. Dehaene, G. Dehaene-Lambertz, Moving along the number line: Operational momentum in nonsymbolic arith- metic. *Percept Psychophys* **69**, 1324–1333 (2007).

97. K. McCrink, K. Wynn, Operational momentum in large-number addition and subtraction by 9-month-olds. *J. Exp. Child Psychol.* **103**, 400–408 (2009).

98. W. Saban, P. Pinheiro-Chagas, S. Borra, R. B. Ivry, Distinct contributions of the cerebellum and basal ganglia to arithmetic procedures. *J. Neurosci.* JN-RM-1482-22 (2023).

99. S. E. Petersen, P. T. Fox, M. I. Posner, M. Mintun, M. E. Raichle, Positron emission tomographic studies of the processing of singe words. *J. Cogn. Neurosci.* **1**, 153–170 (1989).

100. S. E. Petersen, P. T. Fox, M. I. Posner, M. Mintun, M. E. Raichle, Positron emission tomographic studies of the cortical anatomy of single-word processing. *Nature* **331**, 585–589 (1988).

101. S. M. Ravizza, *et al.*, Cerebellar damage produces selective deficits in verbal working memory. *Brain* [Preprint] (2006). Available at: http://dx.doi.org/10.1093/brain/awh685.

102. J. M. Chein, J. A. Fiez, Dissociation of verbal working memory system com- ponents using a delayed serial recall task. *Cerebral Cortex* **11**, 1003–1014 (2001).

103. T. Moberget, E. H. Gullesen, S. Andersson, R. B. Ivry, T. Endestad, Generalized role for the cerebellum in encoding internal models: evidence from semantic processing. *J. Neurosci.* **34**, 2871–2878 (2014).

104. E. Lesage, B. E. Morgan, A. C. Olson, A. S. Meyer, R. C. Miall, Cerebellar rTMS disrupts predictive language processing. *Curr. Biol.* **22**, R794-5 (2012).

105. M. King, S. Bruinsma, R. B. Ivry, No evidence for semantic prediction deficits in individuals with cerebellar degeneration. *Neurobiology of Language* 1–29 (2024).

106. S. T. Piantadosi, F. Hill, Meaning without reference in large language models. *arXiv [cs.CL]* (2022).

107. S. A. Bunge, C. Wendelken, D. Badre, A. D. Wagner, Analogical reasoning and prefrontal cortex: evidence for separable retrieval and integration mechanisms. *Cereb. Cortex* **15**, 239–249 (2005).

108. S. Musker, A. Duchnowski, R. Millière, E. Pavlick, LLMs as models for analogical reasoning. *arXiv [cs.CL]* (2024).

109. F. Van Overwalle, Social and emotional learning in the cerebellum. *Nat. Rev. Neurosci.* **25**, 776–791 (2024).

110. A. Ciricugno, C. Ferrari, L. Battelli, Z. Cattaneo, A chronometric study of the posterior cerebellum's function in emotional processing. *Curr. Biol.* **34**, 1844-1852.e3 (2024).

111. F. Van Overwalle, *et al.*, The role of the cerebellum in reconstructing social action sequences: a pilot study. *Soc. Cogn. Affect. Neurosci.* **14**, 549–558 (2019).

112. S. H. Fatemi, *et al.*, Consensus paper: pathological role of the cerebellum in autism. *Cerebellum* **11**, 777–807 (2012).

113. S. H. Fatemi, T. J. Reutiman, T. D. Folsom, R. W. Sidwell, The role of cerebellar genes in pathology of autism and schizophrenia. *Cerebellum* 1–16 (2007).

114. Z. Chen, D. Whitney, Inferential emotion tracking (IET) reveals the critical role of context in emotion recognition. *Emotion* **22**, 1185–1192 (2022).

115. L. Shahshahani, M. King, C. Nettekoven, R. B. Ivry, J. Diedrichsen, Selective recruitment of the cerebellum evidenced by task-dependent gating of inputs. *Elife* **13** (2024).

116. M. King, L. Shahshahani, R. B. Ivry, J. Diedrichsen, A task-general connectivity model reveals variation in convergence of cortical inputs to functional regions of the cerebellum. *Elife* **12** (2023).

117. S. G. Kim, K. Uğurbil, P. L. Strick, Activation of a cerebellar output nucleus during cognitive processing. *Science* **265**, 949–951 (1994).

118. J. E. Desmond, J. A. Fiez, Neuroimaging studies of the cerebellum: language, learning and memory. *Trends Cogn. Sci.* **2**, 355–362 (1998).

119. A. Chirino-Pérez, *et al.*, Mapping the cerebellar cognitive affective syndrome in patients with chronic cerebellar strokes. *Cerebellum* **21**, 208–218 (2022).

120. M. P. Alexander, S. Gillingham, T. Schweizer, D. T. Stuss, Cognitive impairments due to focal cerebellar injuries in adults. *Cortex* **48**, 980–990 (2012).

121. S. Richter, *et al.*, Cognitive functions in patients with MR-defined chronic focal cerebellar lesions. *J. Neurol.* **254**, 1193–1203 (2007).

122. J. P. Neau, E. Arroyo-Anllo, V. Bonnaud, P. Ingrand, R. Gil, Neuropsychological disturbances in cerebellar infarcts. *Acta Neurol. Scand.* **102**, 363–370 (2000).

123. T. Zhu, *et al.*, Contractions in human cerebellar-cortical manifold structure underlie motor reinforcement learning. *J. Neurosci.* **45**, e2158242025 (2025).

124. W. Heffley, C. Hull, Classical conditioning drives learned reward prediction signals in climbing fibers across the lateral cerebellum. *Elife* **8** (2019).

125. M. J. Wagner, T. H. Kim, J. Savall, M. J. Schnitzer, L. Luo, Cerebellar granule cells encode the expectation of reward. *Nature* **544**, 96–100 (2017).

126. J. E. Trach, Y. Ou, S. D. McDougle, The human cerebellum encodes temporally sensitive reinforcement learning signals. *bioRxiv* (2025).

127. Cisneros, E., Bol, K., Collins, A., Abram, S., Ivry, R. B., Tsay, J. S., No Evidence for the Role of the Cerebellum in Value-based Decision Making in *RLDM Conference*, (2025).

128. Abram, S. J., Tsay, J. S., Wang, T., McDougle, S., Ivry, R., Reinforcement Learning, the Cerebellum, and Agency in *Neural Control of Movement*, (2022).

129. C. M. White, E. C. Snow, A. S. Therrien, Reinforcement motor learning after cerebellar damage is related to state estimation. *Cerebellum* (2023). https://doi.org/10.1007/s12311-023-01615-4.

130. J. Nicholas, *et al.*, The role of the cerebellum in learning to predict reward: Evidence from cerebellar ataxia. *Cerebellum* **23**, 1355–1368 (2024).

131. Z. Cattaneo, *et al.*, New Horizons on non-invasive brain stimulation of the social and affective cerebellum. *Cerebellum* **21**, 482–496 (2022).

132. M. Manto, *et al.*, Consensus paper: Novel directions and next steps of non-invasive brain stimulation of the cerebellum in health and disease. *Cerebellum* **21**, 1092–1122 (2022).

133. M.-E. Bolduc, *et al.*, Spectrum of neurodevelopmental disabilities in children with cerebellar malformations. *Dev. Med. Child Neurol.* **53**, 409–416 (2011).

134. S. S.-H. Wang, A. D. Kloth, A. Badura, The cerebellum, sensitive periods, and autism. *Neuron* **83**, 518–532 (2014).

135. M. C. Granovetter, S. Robert, L. Ettensohn, M. Behrmann, With childhood hemispherectomy, one hemisphere can support-but is suboptimal for-word and face recognition. *Proc. Natl. Acad. Sci. U. S. A.* **119**, e2212936119 (2022).

136. M. Behrmann, D. C. Plaut, Bilateral hemispheric processing of words and faces: evidence from word impairments in prosopagnosia and face impairments in pure alexia. *Cereb. Cortex* **24**, 1102–1118 (2014).

137. S. W. Keele, R. Ivry, Does the Cerebellum Provide a Common Computation for Diverse Tasks? A Timing Hypothesis a. *Ann. N. Y. Acad. Sci.* **608**, 179–211 (1990).

138. H. Fujita, T. Kodama, S. du Lac, Modular output circuits of the fastigial nucleus for diverse motor and nonmotor functions of the cerebellar vermis. *Elife* **9** (2020).

139. J. Beckinghausen, R. V. Sillitoe, Insights into cerebellar development and connectivity. *Neurosci. Lett.* **688**, 2–13 (2019).

140. M. Xie, S. P. Muscinelli, K. Decker Harris, A. Litwin-Kumar, Task-dependent optimal representations for cerebellar learning. *Elife* **12** (2023).